\documentclass[11pt]{article}

\usepackage{hyperref}
\usepackage{amsmath,amsfonts,amssymb}
\hypersetup{dvips,dvipdfm,colorlinks=true,urlcolor=magenta,filecolor=magenta,linktoc=page,citecolor=red,linkcolor=blue,bookmarks=true}
\usepackage{graphicx,epstopdf}
\usepackage{multirow,array}
\begin{document}
	\title{\textbf{Black-Hole singularity and its possible mitigations: Reformulation of Penrose’s Singularity theorem using null Raychaudhuri matrix}}
	\author{Madhukrishna Chakraborty \footnote{chakmadhu1997@gmail.com}~~and~~
		Subenoy Chakraborty\footnote{schakraborty.math@gmail.com (corresponding author)}
		\\ \textit{Department of Mathematics, Jadavpur University, Kolkata - 700032, India}}
	\date{}
	\maketitle
	\begin{abstract}
		In this paper, we introduce a notion of null Raychaudhuri matrix motivated by the matrix representation of tensor fields. The evolution of this matrix gives the matrix form of null Raychaudhuri equation. Using this distinct geometric approach, we have reformulated the original Penrose’s singularity theorem on Black-Hole and have commented on the characteristic of the Raychaudhuri matrix at the singularity. The paper also suggests two possible mitigations for the physical singularity of Schwarzschild black hole via the Wheeler-DeWitt formalism and Bohmian formalism.
	\end{abstract}
	\maketitle
	\section{Introduction}
	Einstein's General theory of Relativity (GR) \cite{Wald:1984rg}, \cite{Weinberg:1972kfs} is the most celebrated theory of gravity which describes a plethora of physical phenomena. Detection of gravitational waves \cite{LIGOScientific:2017vwq}, gravitational red-shift \cite{Wilhelm:2019jou}, gravitational lensing \cite{Bartelmann:2010fz}, perihelion precession of Mercury's orbit \cite{Kraniotis:2003ig} are some of the finest predictions made by Einstein's GR \cite{Wald:1984rg}, \cite{Weinberg:1972kfs}. 
	
	However, the biggest drawback of this theory lies in the existence of singularity \cite{Joshi:2013xoa}. In the context of GR, gravitational singularities are considered where the density becomes infinite both in case of a Black-Hole (BH) \cite{Bhattacharjee:2018kro} as well as in the cosmological context. Even, the simplest solution of GR i.e, the Schwarzschild BH solution is singular. A BH can be described as a region of the space-time manifold where gravity is so strong that even light and other electromagnetic waves fail to escape it \cite{Lan:2023cvz}. To have a proper explanation regarding the space-time singularity in GR, a formal mathematical description is needed. For this, one has to consider the background geometry of GR.

	The background of GR is an $n$-dimensional Lorentzian manifold $M$ equipped with a symmetric metric $g$ which is a non-degenerate $(0,2)$-tensor field assigning to any point $p$ in the smooth manifold $M$, a scalar product $g_{p}$ on the tangent space $T_{p}(M)$ at the point $p$ \cite{Finster:2014dfa}. This scalar product $g_{p}$ is not positive definite but has signature $(-1,+1,+1,...,+1)$.
	
	 Thus, we can visualize GR as a space-time manifold with signature $(n-1,1)$ or equivalently $(1,n-1)$ with a time orientation. Lorentzian manifolds bear some interesting feature like causality and concept of null vectors which make them different from the Riemannian ones. This may be attributed to the fact that the spaces under consideration in studying the Riemannian manifolds are Cauchy complete or equivalently geodesically complete \cite{Lord:2000yt},  \cite{Ehrlich:2006aen}. However, a number of Lorentzian manifolds used as models in GR fail to be geodesically complete. For an arbitrary Lorentzian manifold $(M,g)$, a geodesic $c$ with $\Lambda$ as an affine parameter (by convention) is said to be complete if it can be extended for all values of $\Lambda$ i.e, $-\infty<\Lambda<+\infty$.
	 
	  On the other hand, a past and future inextendible geodesic is said to be incomplete if it cannot be extended to arbitrarily large negative and positive values of the affine parameter. It is to be noted that certain exact solutions to the Einstein's field equations like the extended Schwarzschild solution contain non-space like geodesics which become incomplete upon running into black-holes.
	  
	   In the present work, BH singularity or equivalently the existence of incomplete null geodesic as prescribed by Penrose \cite{Penrose:1964wq} has been attempted from the point of view of Raychaudhuri equation \cite{Raychaudhuri:1953yv}. In the context of BH, the seminal singularity theorem by Penrose gives the key essence that a singularity has the property of geodesic incompleteness which may lucidly expressed in the manner: There are paths of observers through space-time or geodesics that can only be extended for a finite time as measured by a comoving observer \cite{Senovilla:2022vlr}. It is reasonably assumed that at the end of the geodesic, the observer has fallen into a singularity or has encountered some pathology or abnormalities in its path. The laws of physics does not hold here and its a break in the theory itself. 
	   
	 The key argument in Penrose's theorem is the notion of geodesic incompleteness. In his seminal papers \cite{Penrose:1964wq}, \cite{Landsman:2022hrn}, Penrose \cite{Penrose:1964wq}  first
	   gave the formal or rather a geometric definition of what a singularity
	   is. The space-time $(M, g)$ is said to be geodesically complete if all
	   inextendible geodesics are complete.
	   
	   In the definition of completeness the term inextendibility is used often. To understand the concept of inextendibility we first state the definition of an ``extension". A space-time $(M',g')$ is called an extension of a given space-time $(M,g)$ if $(M,g)$ may be isometrically embedded as an open proper subset of $(M',g')$.
	   
	    A space-time having no extensions is said to be inextendible or maximal. A geodesic $c:[a,b]\rightarrow M$ is said to be extendible if there exists another geodesic $\tilde{c}:[\alpha,\beta]\rightarrow M$ with $[a,b]\in [\alpha,\beta]$ and $\tilde{c}|_{[a,b]}=c$, otherwise $c$ is inextendible. A space-time manifold is geodesically complete if all its inextendible geodesics are complete otherwise the space-time manifold is said to be incomplete. 
	    
	    For arbitrary Riemannian manifolds, the metric completeness and geodesic completeness are equivalent as guaranteed by the Hopf-Rinow theorem. However, the picture is quite different in case of Lorentzian manifolds. The completeness of Lorentzian manifolds in terms of causality condition may be stated as follows: every pair of points can be joined by maximizing geodesic $c$ i.e, $L(c)=d(p,q)$. In the Lorentzian case, $d$ fails to be a metric. This is the main reason why unlike Riemannian case, a large number of Lorentzian manifolds used as models in GR fail to be geodesically complete. 
	    
	    Thus the singularity in GR may be attributed due to the background Lorentzian structure which has incomplete time-like/ null geodesics. Even the creator of GR, Einstein himself was worried regarding the presence of singularities in his gravity theory. In this context the singularity theorems in GR are regarded as theoretical accomplishment of GR \cite{Senovilla:2021pdg}, \cite{Senovilla:2014gza}. \\
	
	It was during early 1950's when Raychaudhuri addressed the issue of singularity by formulating an evolution equation for the expansion scalar. This is the famous Raychaudhuri equation (RE) \cite{Kar:2006ms}, \cite{Ehlers:2006aa}, \cite{Kar:2008zz}, \cite{Dadhich:2005qr}  which has a diverse field of application not only in gravity, cosmology and astrophysics but beyond them. Gravitational singularities and the generic features of GR fascinated Raychaudhuri. Although his works were entirely restricted to cosmology, yet it can be immensely applied to BH physics. He first pointed out that singularity is nothing more than an artifact of the symmetries of the matter distribution. When he derived the equations, he considered a time dependent geometry but did not assume homogeneity or isotropy at the outset. He wanted to examine the role of anisotropy (via shear) and rotation (via vorticity) in the formation of a singularity \cite{Dasgupta:2007nr}, \cite{Chakraborty:2023yyz}. \\

	Raychaudhuri equation can be established individually for time-like as well as null geodesic congruence. However, in the present context we are interested in BH. So we invoke the null Raychaudhuri equation and study its corresponding convergence conditions. Actually, it can be shown that in Einstein gravity the RE together with some conditions on matter leads to the focusing of the geodesic congruence in finite value of the affine parameter. This is called the Focusing theorem (FT) which is reasonably believed to be the most important consequence of RE that hints the inevitable existence of singularity in Einstein gravity. 
	
	So FT is the turning point of GR \cite{Chakraborty:2023rgb}, \cite{Ciambelli:2023mir}. Focusing leads to formation of caustic or a congruence singularity \cite{Chakraborty:2023lav}, \cite{Chakraborty:2024wty}. This may not be always a space-time singularity. However, with certain global assumptions, this may sometimes lead to cosmological or BH singularity.  These global assumptions appear in the singularity theorems furnished by Penrose and those have been discussed in greater detail in the subsequent sections of this paper.\\
	
	The novelty of the work lies in the fact that besides stating Penrose's theorem, we have reformulated it from a distinct geometric approach inspired by the matrix representation of tensor fields where we have invoked the notion of conjugate points, Jacobi tensors and identified a null Raychaudhuri matrix whose evolution essentially gives the matrix form of the RE. Focusing theorem may also be restated in matrix analogous form. This formulation is purely geometric which gives the characterization of the null Raychaudhuri matrix at the BH singularity. Moreover, the paper points out the event horizon and BH singularity using a Harmonic oscillator approach of the RE \cite{Chakraborty:2023ork}. The paper also focuses on some possible mitigation approaches. In literature, resolutions of BH singularity have been attempted from Raychaudhuri equation point of view \cite{Blanchette:2021jcw} where they computed the generic corrections to the Raychaudhuri equation in the interior of the Schwarzschild BH, arising from modifications to the algebra inspired by the Generalised Uncertainty Principle (GUP) theories \cite{Blanchette:2021vid}. Further, Rastagoo et. al \cite{Rastgoo:2022mks} studied the interior of the Schwarzschild BH by carrying out a comparative study by Loop Quantum Gravity (LQG) vs GUP using congruences and their associated evolution of expansion scalar i.e the RE. In \cite{Blanchette:2020kkk}, the authors showed a possible resolution of the BH singularity by LQG which leads to the finiteness of expansion and its rate of change in the effective regime in the interior of the Schwarzschild BH.\\
	
	 However, in the present work, we have focused on a two fold mitigation namely, by the quantized null Raychaudhuri equation (canonical quantization) followed by formation of Wheeler-DeWitt (WD) equation and Bohmian formalism involving quantum potential. The formulation of quantized null Raychaudhuri equation finds an immediate application in BH physics, BH thermodynamics and Hawking temperature. An attempt to derive the evolution of quantized time-like geodesic congruence has been made in \cite{Choudhury:2021huy}. The other approach is a semi-classical approach for quantum gravity where the classical geodesics are replaced by quantal bohmian trajectories with a hope that quantum effects which become dominant in the strong gravity regime may alleviate the singularity problem at the classical level. This approach is motivated by the works \cite{Das:2013oda}, \cite{Ali:2014qla}.\\
	
	The layout of the paper is as follows: Section 2 deals with the classical null RE, FT for null geodesics and BH singularity from the point of RE. Section 3 discusses the original Penrose's singularity theorem and the reformulation of it in terms of Raychaudhuri matrix. Section 4 deals with the proposed mitigation schemes and their application in removing the physical singularity of Schwarzschild BH. This section further consists of two subsections namely, 4.1 and 4.2. 4.1 deals with the derivation of quantized null RE and WD formalism while, 4.2 deals with Bohmian formalism and discusses their behavior at the physical singularity of Schwarzschild BH. The paper ends with discussion and future scope in Section 5.
	\section{Classical null Raychaudhuri equation, Focusing theorem and Black-Hole singularity: A Harmonic Oscillator analysis}
	Raychaudhuri equation characterizes the kinematics of flows in a geometrical space. A vector field generates flows and the integral curves of the vector field lead the flow. In Lorentzian space-time manifold, these congruences may be either time-like or null in nature. Historically, the evolution equation for the expansion scalar is known as the RE. However, the evolution of kinematics are in general known as Codazzi-Raychaudhuri equation \cite{Zafiris:1997he}, \cite{Carter:1996wr}. In Lorentzian geometry, Raychaudhuri equation behaves like a geometric identity. The effect of gravity comes through the Ricci tensor term projected along the geodesic \cite{Chakraborty:2019vki}. In case of null geodesic congruence, the tangent vector field here is a null vector $k^{a}$.  We choose the geodesics to be affinely parametrized by the parameter $\Lambda$. Thus we can write, $dx^{a}=k^{a}d\Lambda$. Let $\eta^{a}$ denotes the deviation vector. Thus one may write,
	\begin{equation}
		k^{a}k_{a}=0,~k^{b}\nabla_{b}k^{a}=0,~\eta^{b}\nabla_{b}k^{a}=k^{b}\nabla_{b}\eta^{a},~k^{a}\eta_{a}=0
	\end{equation}
Since, $k^{a}$ is null, so $h_{ab}=g_{_{ab}}+k_{a}k_{b}$ fails to work as the transverse metric. Therefore, it is necessary to introduce another null vector $N^{a}$ so that $k^{a}N_{a}=-1$, using arbitrariness of normalization of a null vector. Thus, a purely transverse and effectively two dimensional metric may be written as
\begin{equation}
	h_{ab}=g_{ab}+k_{a}N_{b}+N_{a}k_{b}
\end{equation}
 and we have, $h_{ab}k^{b}=h_{ab}N^{B}=0$, $h^{a}_{a}=2$, $h^{a}_{c}h^{c}_{b}=h^{a}_{b}$. The spatial tensor field is given by
 \begin{equation}
 	\tilde{B_{ab}}=\dfrac{1}{2}\Theta h_{ab}+\sigma_{ab}+\omega_{ab}
 \end{equation} with $\Theta=\tilde{B}^{a}_{a}$ being the expansion scalar, $\sigma_{ab}=\tilde{B_{(ab)}}-\dfrac{1}{2}\Theta h_{ab}$ is the shear tensor and $\omega_{ab}=\tilde{B}_{[ab]}$ is the rotation tensor. It follows that $\Theta=\nabla_{a}k^{a}$ and the corresponding null RE is given by 
\begin{equation}
	\dfrac{d\Theta}{d\Lambda}=-\dfrac{\Theta^{2}}{2}-\sigma_{ab}\sigma^{ab}+\omega_{ab}\omega^{ab}-R_{cd}k^{c}k^{d}
\end{equation}
This is the evolution equation for expansion scalar known as RE for null geodesic congruence or simply null RE. Using Einstein's field equations, one has
\begin{equation}
	R_{ab}k^{a}k^{b}=8\pi T_{ab}k^{a}k^{b}
\end{equation} as for null geodesic congruence we have $k^{a}k_{a}=0$. If the Null Energy Condition (NEC) is satisfied i.e, $R_{ab}k^{a}k^{b}\geq0$ then from Einstein's equation it follows that $T_{ab}k^{a}k^{b}\geq0$ for all $k^{a}$. NEC follows from the assumption of Strong Energy Condition (SEC). For hyper-surface orthogonal null geodesic congruence we have $\omega_{ab}=0$ by virtue of Frobenius theorem. This yields,
\begin{equation}
	\dfrac{d\Theta}{d\Lambda}\leq-\dfrac{\Theta^{2}}{2}
\end{equation} which upon integration yields
\begin{equation}
	\dfrac{1}{\Theta(\Lambda)}\geq \dfrac{1}{\Theta_{0}}+\dfrac{\Lambda}{2}
\end{equation} where $\Theta_{0}=\Theta(\Lambda=0)$. Therefore, for an initially converging hyper-surface orthogonal congruence of null geodesics, focusing is inevitable for a finite value of the affine parameter given by $\Lambda\leq\dfrac{2}{|\Theta_{0}|}$. This is the null analogue of the FT and the corresponding condition for focusing i.e, $R_{ab}k^{a}k^{b}\geq0$ is known as the Convergence Condition (CC). 

It is to be noted that focusing alone does not lead to formation of singularity. But the converse is true. This means, if there is a singularity in the space-time manifold, geodesics will inevitably focus at the singularity. This motivates us to get a criteria for the possible avoidance of singularity. Since a singularity implies focusing, so if we can avoid the focusing by any means we can possibly avoid the formation of singularity. In other words, if focusing/CC is violated then the geodesics can be extended for infinitely large positive and negative values of the affine parameter and there is no singularity. 

Mathematically, RE can be termed as Riccati equation and become analogous to the differential equation of a Harmonic Oscillator with frequency varying with affine parameter as follows
\begin{equation}
	\dfrac{d^{2}Y}{d\Lambda^{2}}+\omega_{0}^{2}Y=0
\end{equation} where 
\begin{equation}
	\Theta=2\dfrac{d}{d\Lambda}\ln Y
\end{equation} and 
\begin{equation}
	\omega_{0}^{2}=\dfrac{1}{2}(\tilde{R}+2\sigma^{2}-2\omega^{2})
\end{equation} where $\tilde{R}=R_{ab}k^{a}k^{b}$ is called the Raychaudhuri scalar. $\Theta$ may be defined as the derivative of $S$, the geometric entropy or an effective geodesic deviation. So, one may identify $S=\ln Y$. The expansion $\Theta$ gives the rate of change of volume of the transverse subspace of the congruence or bundle of geodesics. Thus $\Theta \rightarrow-\infty$ (expansion approaching negative infinity) implies a complete convergence while $\Theta \rightarrow +\infty$ (expansion approaching positive infinity) implies a complete divergence of the bundle of geodesics. One may restate the convergence condition as follows:
\begin{enumerate}
	\item Initially, $Y>0$ but $\dfrac{dY}{d\Lambda}<0$.
	\item Subsequently, $Y=0$ at a finite proper time in order to have negatively infinite expansion. 
	\end{enumerate} From the above relation between the variables $\Theta$ and $Y$, i.e, $\Theta=\dfrac{2}{Y}\dfrac{dY}{d\Lambda}$, it is clear that there should be an initial negative expansion i.e, ($\Theta(\Lambda)<0$ and subsequently $\Theta\rightarrow-\infty$ as $Y\rightarrow0$ at a finite value of the affine parameter $\Lambda$). The CC is essentially the condition to get zeros of $Y$. In the theory of differential equation, we have the Strum Comparison theorem which shows that existence of zeros in $Y$ at finite value of the affine parameter. The required condition is given by
\begin{equation}
(R_{ab}k^{a}k^{b}+2\sigma^{2}-2\omega^{2})\geq0
\end{equation} where $2\sigma^{2}=\sigma_{ab}\sigma^{ab}$ and $2\omega^{2}=\omega_{ab}\omega^{ab}$. The above inequality shows that if $R_{ab}k^{a}k^{b}\geq0$ then shear (anisotropy) favors convergence while rotation (vorticity) defies it. So we name $R_{c}=\tilde{R}+2\sigma^{2}$ as the convergence scalar. This motivates us to examine BH singularity via the Harmonic oscillator form of the RE. As an example and for the sake of simplicity we consider the Schwarzschild metric given by
\begin{equation}
	ds^{2}=-\left(1-\dfrac{2GM}{r}\right)dt^{2}+\left(1-\dfrac{2GM}{r}\right)^{-1}dr^{2}+r^{2}d\Omega^{2}
\end{equation} where $d\Omega^{2}$ is the metric on unit 2-sphere.
We know that Schwarzschild metric is a vacuum solution of the Einstein's field equations. In GR, a vacuum solution is a Lorentzian manifold whose Einstein tensor and hence Ricci tensor vanishes identically i.e, $\tilde{R}=0$. Thus, for hyper-surface orthogonal congruence of null geodesics we have $\omega_{0}^{2}=\tilde{R}+2\sigma^{2}-\omega^{2}=2\sigma^{2}$. Using the dynamics of classical Schwarzschild metric and using the definition of anisotropy scalar $\sigma^{2}$ we have
\begin{equation}
	\omega_{0}^{2}=2\sigma^{2}=\dfrac{4}{3r^{3}}\dfrac{(3GM-r)^{2}}{(2GM-r)}
\end{equation} Now we identify the physical singularity, geometric singularity and event horizon for Schwarzschild BH using this approach.
\begin{itemize}
	\item $R_{c}$, the convergence scalar is positive only when $r<2GM$. Thus the region $r<2GM$ is favorable for convergence of null geodesics. This gives the possibility that singularity may occur in this region. As $r\rightarrow0$, at some finite value of the affine parameter $R_{c}$ will dominate over $\omega^{2}$ so that $R_{c}-2\omega^{2}\geq0$ i.e, CC will hold hinting the fact that $r=0$ is a singularity.
	\item We also find that for $r>GM$, $R_{c}<0$. Thus, there is no singularity in the region $r>2GM$ as focusing is violated in this region.
	\item $r=2GM$ acts as a boundary to distinguish the singular and non-singular regions and this is known as the event horizon.
	\item The expression for $R_{c}$ shows that $r=0$ and $r=2GM$ are the points where $R_{c}$ diverges. These are therefore, the points of singularity. The singularity at $r=2GM$ is a coordinate singularity that arises due to the bad choice of coordinates. This singularity can be removed with suitable choice of coordinate namely the Eddington-Finkelstein coordinate. However, the singularity at $r=0$ is the physical singularity or the BH singularity whose existence is guaranteed by the RE.
	\item The Harmonic Oscillator approach shows that CC is associated with the frequency of the oscillator. Thus, if RE can be expressed as the evolution equation for a real harmonic oscillator with varying frequency then it is inevitable to have a singularity. In the expression for $R_{c}$, if $r=o(\delta)$, then as $r\rightarrow0$, $\omega_{0}^{2}$ varies largely with $o(\delta)^{3}$ as compared to $r\rightarrow 2GM$ where the frequency varies with $o(\delta)$.
	
	 Though focusing can occur in Minkowski spacetime
	(complete vacuum), a singularity cannot occur in a vacuum. The interior geometry of a massive star develops a singularity due to collapse. BH singularity is such a case.
	\end{itemize}
	\section{Penrose's 1965 Singularity theorem and its reformulation using the notion of null Raychaudhuri matrix}
	The acclaimed Singularity theorems are often quoted as one of the greatest theoretical accomplishments in General relativity, Lorentzian geometry and Mathematical physics. Penrose in 1965 brought this revolutionary idea of geodesic incompleteness and linked it with singularity of space-time. He used the concept of a Cauchy hyper-surface (and thereby global hyperbolicity) and more importantly the notion of closed trapped surfaces \cite{Asgari:2024lko}. The original 1965 theorem states that \cite{Penrose:1964wq}
	\textit{``Let $(M,g)$ be a space-time such that
	\begin{enumerate}
		\item $Ric(X,X)\geq0$ for all null vectors $X$ i.e, NEC holds.
		\item There is a non-compact Cauchy surface $S$.
		\item There is an achronal closed future trapped surface $P$.
\end{enumerate} ~~~~~Then $M$ is future null geodesically incomplete."} The key argument in the proof of the singularity theorems is to show that certain curvature conditions force every complete null geodesic to have a pair of conjugate points. In order to define the notion of conjugate points, lets introduce the definition of Jacobi tensors for, the Jacobi tensors are a convenient way of studying conjugate points. Penrose's theorem is about the incomplete null geodesic/ space-times which are null geodesically incomplete.

 In case of Lorentzian manifold, the whole causality theory is based on the fundamental result that states: ``In any normal neighborhood of a point $p$ the exponential map defines a light cone on the manifold such that null geodesics are precisely on that light cone and any other causal curve lies in the interior of the cone from the point that it fails to be a null geodesic. Although, one has the examples of cosmological model of expanding universe or Schwarzschild metric it remains quite subtle to establish a general notion of singularity in GR. \\

 If we define singularity to be a point in the space-time manifold where curvature becomes infinite/unbounded, then it is incompatible with the dynamics of space-time in GR \cite{Racz:2022zis}. 
 But we can talk about a point in the space-time if we can solve the field equations around it to find the corresponding manifold and metric structure. Therefore, singularity should be viewed as some kind of singular boundary point of space-time. In that case, the question that still remains is how to detect the occurrence of a singularity from within space-time. It was first established in Penrose's work that the two concepts: Singularities and geodesic incompleteness were aligned \cite{Vargas-Serdio:2020qpu}. After this, we call a space-time singular if it contains an incomplete causal geodesic.\\
 
  This can be elaborated by an example- A future incomplete causal geodesic corresponds to a freely falling observer or to a light ray that suddenly ends its existence, while in the past it corresponds to a light ray that suddenly comes into existence out of nowhere. Both of these situations are more objectionable than divergence of the curvature term. Thus, the general point of view has become to regard causal geodesic completeness as a minimal condition for the existence of singularity free space-time. Null geodesic completeness implies inextendibility (definition of extension and inextendibility have been discussed in the Introduction). Thus, extendibility can be a source of incompleteness and needs to be excluded. Penrose's theorem covers the gravitational collapse scenario. \\
  
  To understand the structures that build the theorem we first introduce the notion of conjugate points and the relevant information on conjugate points is contained in the concept of Jacobi tensors \cite{glg}. Given a geodesic $c$, define $N(c(t))$ to be the $(n-1)$-dimensional subspace of Jacobi fields vanishing at a given point and taking values in $\dot{c(t)}^{\perp}:=\{v\in T_{c(t)}M,~<v,\dot{c(t)}>=0\}$. A Jacobi field is a vector field $J$ along a geodesic $c$ that satisfies the Jacobi equation
\begin{equation}
	\ddot{J}+R(J,\dot{c})\dot{c}=0,
\end{equation} $\ddot{J}$ denotes the covariant derivative along $c$. In the context of the singularity theorems, focusing effect on geodesic is exerted by certain curvature conditions. The notion of conjugate points is very much essential for the focusing of geodesics. Two points $c(a)$ and $c(b)$ are called conjugate points along $c$ if there exists a non-trivial Jacobi field vanishing at $a$ and $b$. To properly deal with the null case (since we are interested in BH) where $\dot{c(t)}=\dot{c(t)}^{\perp}$ we use the quotients $[\dot{c(t)}]^{\perp}:=\bigcup_{t}[\dot{c(t)}]^{\perp}$ respectively.\\

 It is to be noted that unlike the time-like case where $[\dot{c(t)}]^{\perp}$ coincides with $\dot{c(t)}^{\perp}$, in null case the dimension of $[\dot{c(t)}]^{\perp}$ is $n-2$. The restriction of the metric $g|_{[\dot{c(t)}]^{\perp}}$ is positive definite. We now introduce a $(1,1)$ tensor field, known as Jacobi tensor along the null geodesic $c$ by $[A]:[\dot{c}]^{\perp}\rightarrow [\dot{c}]^{\perp}$ which gives the tensor Jacobi equation $[\ddot{A}]+[R][A]=0$, with $[R]:[v]\rightarrow [R(v,\dot{c})\dot{c}]$ is the tidal force operator and an extra condition is given by $ker[A(t)]\cap ker[\dot{A(t)}]=\{[\dot{c(t)}]\}$ for all $t\in[a,b]$. $[R][A]$ is an endomorphism on $c(t)$ defined by 
\begin{equation}
	RA(t)(v)=R(A(t)v,\dot{c})\dot{c}
\end{equation} A Jacobi tensor $A$ is said to be a Lagrange tensor field if $A'^* A-A^*A'=0$ i.e, Wronskian ($[A]^{*},[A]$=0) for all $t\in [a,b]$. We now derive the matrix analogue of the null RE using the matrix representation of the tensor class $[A]$, remembering the fact that we work modulo $\dot{c}$ in $N(c(t))$ and dim$(N(c(t)))=n-2$. The adjoint $[A]^{*}$ of $[A]$ is defined by 
\begin{equation}
	g([A]w,v)=g([A]^{*}v,w)
\end{equation}
For the matrix representation we consider an orthonormal basis $\{E_{1},E_{2},...,E_{n}\}$ along $c$ with $E_{n-1}=[c]^{\perp}$ and $E_{n}[\dot{c}]$ as $[c]^{\perp}$ and $[\dot{c}]$ both are null for any null vector $c$. Then $N(c(t))=span\{E_{1},E_{2},...,E_{n-2}\}$. Therefore, any Jacobi field in $N(c(t))$ can be expressed as $J=\sum_{i=1}^{n-2}J_{i}E_{i}$. Thus $J$ may be expressed as a column vector  \begin{align}
	J &= \begin{bmatrix}
		J_{1} \\
		J_{2} \\
		\vdots \\
		J_{n-2}
	\end{bmatrix}.
\end{align}
Using this representation, let $J_{i}=J_{i}(t)$ be the column vector corresponding to the Jacobi field along $c$ which satisfies $J(t_{0})=0,~\dot{J(t_{0})}=E_{i}(t_{0})$. Let $A(t)=[J_{1}(t),~J_{2}(t),...,J_{n-2}(t)]$ be the $(n-2)\times (n-2)$ matrix with $J_{i}(t)$ in the $i$-th column. This matrix $A(t)$ is a representation of a Lagrange tensor field along $c$. Using the same basis, we can define the adjoint of matrix $A(t)$ by $A^{*}(t)$. Conjugate points of $c(t_{0})$ along $c$ are exactly those points where $det(A(t))=0$. Hence, $det(A(t))$ has isolated zeros on the interval $[a,b]$. Further, one may calculate the multiplicity of a conjugate point $t=t_{1}$ to $t_{0}$ along $c$ which is just the nullity of $A(t_{1}):N(c(t_{1}))\rightarrow N(c(t_{1}))$. \\

Now we define a new matrix $B$ using the matrix $A$ by $B=\dot{A}A^{-1}$, where $A$ is a Jacobi tensor matrix along a null geodesic $c$ and $B$ is defined at points where $A^{-1}$ is defined. We now express the kinematic variables in terms of the matrix $B$.
\begin{enumerate}
	\item We define expansion $\Theta=tr(B)=(detA)^{-1}(\dot{detA})$ (using result of matrix algebra).
	\item Vorticity tensor $\omega=\dfrac{1}{2}(B-B^{*})$.
	\item Shear tensor $\sigma=\dfrac{1}{2}(B+B^{*})-\dfrac{\Theta}{n-2}E_{n-2}$, where $E_{n-2}$ is the identity matrix of order $(n-2)$.
\end{enumerate} So we have $B=\dfrac{1}{2}(B+B^{*})+\dfrac{1}{2}(B-B^{*})=\sigma+\omega+\dfrac{\Theta}{n-2}E_{n-2}$. Using $(det{A}^{-1})=-A^{-1}\dot{A}A^{-1}$ we get, $\dot{B}=-R-B^{2}$. This is the evolution equation of the matrix RE and is analogous to a special type of Riccati equation. Using $\Theta=tr(B)$, $B=\sigma+\omega+\dfrac{\Theta}{n-2}E_{n-2}$ where $E_{n-2}$ is the $(n-2)\times(n-2)$ identity matrix. Consequently, we can write
\begin{equation}
	\dot{\Theta}=tr(\dot{B})=tr(-R-B^{2})=-tr(R)-tr(B^{2})
\end{equation} (using the linearity of trace). Finally, by using the expression for $B$, $tr(E_{n-2})=n-2$, $tr(\omega)=0$, $tr(\sigma)=0$ and $tr(\sigma \omega)=0$. Now we calculate $tr(R)$ as follows:
Let $V(c)$ be the geometric realization for $N(c(t))$ already constructed. Let $\{E_{1},E_{2},...,E_{n-2}\}$ be an orthonormal basis for $N(c(t))$ at every point of $c(t)$. Next, we extend $\{E_{1}, E_{2},...,E_{n-2}\}$ to an orthonormal basis $\{E_{1},E_{2},...,E_{n}\}$ along $c$, where $E_{n}$ is time-like and $\dot{c}=\dfrac{E_{n-1}+E_{n}}{\sqrt{2}}$. Then we have 
\begin{eqnarray}
	g(R(E_{n-1}, \dot{c})\dot{c},E_{n-1})-g(R(E_{n},\dot{c})\dot{c}, E_{n})\\
	=\dfrac{1}{2}\left(g(R(E_{n-1}, E_{n})E_{n},E_{n-1}\right)-g(R(E_{n}, E_{n-1})E_{n-1},E_{n})=0\nonumber
\end{eqnarray} where the basic properties of curvature have been used. Consequently, we get \begin{eqnarray}
tr(R)=\sum_{i=1}^{n-2}g(R(E_{i},\dot{c})\dot{c}, E_{i})\\
~~~~~~~~~~~~~~~~~~~~~~~~~~~=\sum_{i=1}^{n}g(E_{i},E_{i})g (R(E_{i},\dot{c})\dot{c}, E_{i})\\ \nonumber
=Ric(\dot{c},\dot{c}).\nonumber
\end{eqnarray} Thus the null RE from the evolution equation of matrix $B$ turns out to be
\begin{equation}
	\dot{\Theta}=-Ric(\dot{c},\dot{c})-tr(\sigma^{2})-tr(\omega^{2})-\dfrac{\Theta^{2}}{n-2}
\end{equation} Since the evolution of the matrix $B$ gives the null RE so we name $B$ as \textbf{\textit{null Raychaudhuri matrix}}. If $A$ is a lagrange tensor field i.e, $A^{*}\dot{A}-\dot{A}A^{*}=0$ then $B=B^{*}$ and hence $\omega=0$. So we get the vorticity free null RE as 
\begin{equation}
		\dot{\Theta}=-Ric(\dot{c},\dot{c})-tr(\sigma^{2})-\dfrac{\Theta^{2}}{n-2}
\end{equation} This approach has a physical significance as it may be treated within the frame work of conjugate points and oscillation theory in ordinary differential equations. In this approach, the RE can be transformed by a change in variable to the differential equation already mentioned in the previous section
\begin{equation}
	\ddot{y}+G(\Lambda)\dot{y}=0
\end{equation} where 
\begin{eqnarray}
	G(\Lambda)=\dfrac{1}{n-2}\left(Ric(\dot{c},\dot{c})+2\sigma^{2}\right)\\ \nonumber
	=\dfrac{1}{2}\left(Ric(\dot{c},\dot{c})+2\sigma^{2}\right)
\end{eqnarray} as $n=4$ in the present context. Here over dot denotes differentiation w.r.t affine parameter $\Lambda$.\\

 The matrix form of FT can be expressed as follows: ``Let $c:[a,b]\rightarrow M$ be an inextendible null geodesic satisfying $Ric(\dot{c},\dot{c})\geq0$ (NEC) for all $\Lambda\in J=[a,b]$. Let $A$ be the Lagrange tensor field along $c$. Suppose that the expansion $\Theta=tr(\dot{A}A^{-1})$ has initially a negative value at $\Lambda_{1}\in J$. Then $det(A)=0$ for some $\Lambda$ in the interval from $\Lambda_{1}$ to $\Lambda_{1}-\dfrac{n-2}{\Theta_{1}}$ provided that $\Lambda\in J$."
 
 \textbf{Remark:} If a singularity exists then at least one of the eigen values is not finite. Suppose a bundle of null geodesics start at $\Lambda=a$ and meet again at $\Lambda=b$. This implies, one of the eigen values is zero at $\Lambda=a$ and the conjugate point corresponds to the next value of $\Lambda>a$ at which one of the eigen values is zero. The differential equation in $B$ is a special type of Riccati equation. But the differential equation in $A$ i.e, $\ddot{A}+RA=0$ implies $(\ddot{tr(A)})+tr(R)tr(A)=0$ (using linearity of trace and $tr(AB)=tr(A)tr(B)$). Thus if, $tr(R)\geq0$ then $tr(A)$ has oscillatory behavior.
 
  In order to restate the original Penrose's theorem, we introduce few more notions. For two points $p,q\in M$ we define $p<<q$ if there is a future directed null curve from $p$ to $q$. $p\leq q$ means there exists a future directed causal curve from $p$ to $q$. Based on this we may define the chronological and causal future as the following sets \cite{Steinbauer:2022hvq}
\begin{eqnarray}
~~~	I^{+}(p)=\{q\in M|~p<<q\}\nonumber\\
J^{+}(p)=\{q\in M|~~~p\leq q\}\nonumber
\end{eqnarray}
With the same analogy we may define chronological and causal past by $I^{-}(p)$ and $J^{-}(p)$. If $c:[a,b]\rightarrow M$ (a sufficiently smooth curve), we define its length by $L(c)=\int_{[a,b]}\sqrt{|<\dot{c(t)},~\dot{c(t)}>|}d\Lambda$. The Lorentzian distance on $M\times M$ is defined by \begin{equation}
d(p,q)=
\begin{cases}
	sup~L(c) & \text{if } q \in J^{+}(p)\\
	0 & else
\end{cases}
\end{equation}
On the contrary, the Riemannian distance is defined by $inf~L(c)$. \\

One may check that the metric in Lorentzian case fails to be symmetric and it satisfies the reverse triangle inequality $d(p,q)\geq d(p,r)+d(r,q)~ \forall ~p\leq r \leq q$ i.e, detouring makes curves longer rather than shorter. Also the supremum is not finite if the causality of the space-time behaves badly e.g if there are closed time-like curves. The completeness of Riemannian manifolds in terms of causality condition may be stated as follows: every pair of points can be joined by a maximizing geodesic $c$ that is $L(c)=d(p,q)$. In the Lorentzian case, $d$ fails to be a metric. It fails to be upper semicontinuous in general while it is lower semicontinuous where it is finite. This is the reason why unlike Riemannian case, a large number of Lorentzian manifolds used as models in General Relativity fail to be geodesically complete.\\

 There is however a class of space-times where $d$ is finite and continuous namely \textit{globally hyperbolic} space-time. By \textit{Avez-Seifert} theorem, \textit{globally hyperbolic} space-times are causal geodesically connected. This means, any pair of points $p\leq q$ can be joined by a causal geodesic say, $c$ that is maximizing $L(c)=d(p,q)$. A future directed non space-like curve $c$ from $p$ to $q$ is said to be maximal if $L(c)=d(p,q)$. The analogy of geodesics in a complete Riemannian manifold is there for Lorentzian manifold provided it is globally hyperbolic. A space-time is globally hyperbolic if it admits a Cauchy surface.  More precisely, a space-time is said to be singular if it contains an incomplete causal geodesic. However in Lorentzian manifold, time-like geodesic completeness, null geodesic completeness and space-like geodesic completeness all are logically inequivalent. In the context of singularity or geodesic incompleteness the notion of \textit{Achronal Set} is very important. It is a subset $S$ of a Lorentzian manifold $(M,g)$ if none of its points is in the chronological future of any other or there are no such points $p,q\in S$ such that $q\in I^{+}(p)$ or $I^{+}(S)\cap S=\Phi$ (empty) i.e, no two points are time-like related. Readers may find the definitions in greater detail in \cite{glg}, \cite{Steinbauer:2022hvq}. Further, one may define \textit{Future (Past) Horismos} for any set $S$ as \cite{glg}, \cite{Steinbauer:2022hvq}
\begin{eqnarray}
E^{+}(S)=J^{+}(S)-I^{+}(S)\nonumber\\
E^{-}(S)=J^{-}(S)-I^{-}(S)\nonumber
\end{eqnarray}
An \textit{Achronal set} $S$ is future (past) trapped if $E^{+}(S)~(E^{-}(S))$ is compact. In topology, a space is compact if it is closed and bounded. In the context of manifolds, compactness implies that the manifold is finite in size and does not extend infinitely. Let $(M,g)$ be a future null complete space-time where NEC is satisfied. Then any achronal closed future trapped surface is also a future trapped surface. Clearly, $S\subset E^{+}(S)$. For $E^{+}(S)$ to be achronal and compact, $S$ itself needs to be compact. $E^{+}(S)\subseteq \partial J^{+}(S)=\partial I^{+}(S)$ is always true. However the equality occurs both locally and globally in case of \textit{globally hyperbolic} space-times. This may be clearly understood choosing $S$ to be a singleton set $S=\{s\}$. Then $J^{+}(s)$ is the future light cone including the surface while $I^{+}(s)$ contains only the interior of the future light cone. Thus, $E^{+}(s)$ may be identified as the null surface of the cone.

Thus we have everything in hand to restate Penrose's theorem. The restated version can be read as:\\

\textit{Let $(M,g)$ be an $n$-dimensional manifold (strongly causal) and $c:[a,b]\rightarrow M$ be a null geodesic. Then there exists an $(n-2)$ dimensional subspace of space-like vectors $N(c(\Lambda))$ consisting of space-like vectors with the following property:
\begin{enumerate}
	\item There exists no $p,q\in N(c(\Lambda))$ such that $q\in I^{+}(p)$ or $I^{+}(N(c(\Lambda))\cap N(c(\Lambda))=\Phi$.
	\item $E^{+}(N(c(\Lambda)))$ is closed and compact.
	\item $tr(A)$ has oscillatory behavior where $ A_{(n-2)\times (n-2)}$ is the matrix representation of Lagrange tensor field along $c$ w.r.t orthonormal basis $\{ E_{1},E_{2},...,E_{n}\}$ with $ E_{n-1}=c^{\perp}$ and $E_{n}=\dot{c}$
\end{enumerate}
so that $A$ will be a singular matrix in some finite value of the parameter $\Lambda$ and consequently the \textit{\textbf{null Raychaudhuri matrix is undefined there in}}.}\\

\textbf{Remark:} The above form of the Penrose's theorem gives a complete mathematical description of incomplete null geodesic or BH singularity. Raychaudhuri equation together with some global assumptions are the key ingredients of original Penrose's theorem. Similarly in this representation, the matrix analogue of null RE and some other mathematical version of the physical assumptions (mostly in set theoretic approach) are used to restate the Penrose's theorem in a different way. \\

The advantage of this reformulation lies in the fact that if we can identify the RE with its matrix analogue and introduce the corresponding null Raychaudhuri matrix, then the characterization of the later at the BH singularity can be understood. In other words, a mathematical condition of BH singularity can be realized in the way that at the BH singularity the null Raychaudhuri matrix (whose evolution gives the null RE) becomes undefined. Just like the way, Penrose associated the notion of BH singularity with null geodesic incompleteness, we can associate BH singularity with the divergence/blowing of null Raychaudhuri matrix. Based on the above discussion we can infer that BH singularity, null geodesic incompleteness and divergence of null Raychaudhuri matrix are all equivalent.
\section{Mitigation of physical singularity of Schwarzschild Black-Hole}
In the previous section, we showed the existence of BH singularity via RE or Raychaudhuri matrix. At the singularity, no physics can be defined. Thus, some possible mitigation should be adopted. Motivated by the fact that quantum effects which become dominant in strong gravity regime may mitigate the singularity problem that persists at the classical level, we adopt two quantum mechanical approaches for the possible resolution of physical singularity at $r=0$ of Schwarzschild BH, namely WD formalism and Bohmian formalism \cite{Chakraborty:2023voy}. 

Although there is no universal theory of quantum
gravity, there are two major approaches for
formulating a quantum theory of gravity namely canonical quantization and path integral formulation.
In canonical quantization , the operator version of the Hamiltonian
constraint (known as Wheeler-Dewitt (WD) equation) is
a second order hyperbolic functional differential equation
and its solution is known as the wave function of the universe. However, even in simple minisuperspace models it is hard to find a solution of the WD equation \cite{Wheeler:1968iap}, \cite{Jalalzadeh:2016gqs}. Also
there is an ambiguity in operator ordering and to
know the initial conditions of the universe to have a well
defined wave function.\\

 However an important feature of
the Hamiltonian in the operator version is that it admits
a self adjoint extension in a general sense. As a result, the
conservation of probability is ensured. On the other hand,
the path integral formulation is more favourable due to
some definite proposals for the sum over histories. Behavior of singularity at the quantum domain may be revealed by examining the RE in quantum settings. In other words, a quantum version of RE or quantum replica of classical geodesics might be of some help. However, the quantization scheme of RE bear some issues.
 RE is essentially an identity in the Lorentzian geometry so naturally RE cannot be obtained from a variational
principle as the equation of motion for a geodesic congruence. However, when we express the curvature scalar in
terms of Einstein’s field equations (or modified field equations) then RE is no longer an identity, rather there is
some meaning as Lagrangian formulation. \\

Quantum techniques adopted in this paper are canonical quantization and Bohmian formalism. In order to proceed with the canonical quantization we consider the Lagrangian formulation of RE. For this we consider a system with $d$ degrees of freedom. The system is described by $d$ second order differential equation of the form
\begin{equation}
	F_{i}(\Lambda,x_{j},\dot{x_{j}},\ddot{x_{j}})=0\label{eq27}
\end{equation} where over dot denotes derivatives w.r.t the affine parameter $\Lambda$ and $j=1,2,...,d$. Helmholtz conditions are the necessary and sufficient conditions which must be satisfied by (\ref{eq27})
to be the Euler-Lagrange equation corresponding to a Lagrangian $L(\Lambda,x_{j},\dot{x_{j}})$. The Helmholtz conditions are given by \cite{Davis:1928}-\cite{Nigam:2016}
\begin{eqnarray}
	\dfrac{\partial F_{i}}{\partial\ddot{x_{j}}}=\dfrac{\partial F_{j}}{\partial\ddot{x_{i}}} \label{eq28}\\
\dfrac{\partial F_{i}}{\partial x_{j}}-\dfrac{\partial F_{j}}{\partial x_{i}}=\dfrac{1}{2}\dfrac{d}{d\Lambda}\left(\dfrac{\partial F_{i}}{\partial\ddot{x_{j}}}-\dfrac{\partial F_{j}}{\partial \ddot{x_{i}}}\right) \label{eq29}\\
		\dfrac{\partial F_{i}}{\partial\dot{ x_{j}}}+\dfrac{\partial F_{j}}{\partial \dot{x_{i}}}=2\dfrac{d}{d\Lambda}\left(\dfrac{\partial F_{j}}{\partial \ddot{x_{i}}}\right)\label{eq30}
\end{eqnarray} for all $i,j=1,2,...,d$. We can utilize these conditions in the context of the representation of a geodesic congruence as a dynamical system. We consider a hyper-surface orthogonal null geodesic congruence in an $(n+1)$-dimensional space-time. If we consider the congruence as a dynamical system, one can define the dynamical degree of freedom as \cite{Alsaleh:2017ozf}
\begin{equation}
	\rho(\Lambda)=\sqrt{\eta}
\end{equation} where $\Lambda$ is the affine parameter, $\eta=det(\eta_{ab})$ and $\eta_{ab}$ is the induced metric on the $n$-dimensional hyper-surface. The dynamical evolution of $\eta$ is given by
\begin{equation}
	\dfrac{1}{\sqrt{\eta}}\dfrac{d\sqrt{\eta}}{d\Lambda}=\Theta
\end{equation} where $\Theta=\nabla_{a}k^{a}$ is the expansion scalar of the null geodesic congruence. Thus we have, 
\begin{equation}
	\dfrac{1}{\rho}\dfrac{d\rho}{d\Lambda}=\Theta
\end{equation} The RE in terms of the transformed variable $\rho$ is a second order differential equation given by
\begin{equation}
	\dfrac{\ddot{\rho}}{\rho}+\left(\dfrac{2-n}{n-1}\right)\left(\dfrac{\dot{\rho}}{\rho}\right)^{2}+2\sigma^{2}+\tilde{R}=0\label{eq34}
\end{equation} We now compare (\ref{eq34}) with (\ref{eq27}) to get
\begin{equation}
	F=	\dfrac{\ddot{\rho}}{\rho}+\left(\dfrac{2-n}{n-1}\right)\left(\dfrac{\dot{\rho}}{\rho}\right)^{2}+2\sigma^{2}+\tilde{R} \label{eq35}
\end{equation}
We find that Helmholtz conditions (\ref{eq28}) and (\ref{eq29}) are trivially satisfied for the above $F$ given by (\ref{eq35}) and (\ref{eq30}) is satisfied only if $n=3$. So, a Lagrangian formulation with this $F$ is possible only for $n=3$ i.e, only in $(3+1)$ dimensional space-time. To achieve a Lagrangian formulation for the $(n+1)$ dimensional manifold we write
\begin{equation}
	\tilde{F}=\rho^{\alpha}F=\rho^{\alpha-1}\ddot{\rho}+\left(\dfrac{2-n}{n-1}\right)\rho^{\alpha-2}\dot{\rho}^{2}+(2\sigma^{2}+\tilde{R})\rho^{\alpha}
\end{equation} then (\ref{eq30}) gives $\alpha=\dfrac{3-n}{n-1}$. So if we multiply $F$ by $\rho^{\left(\dfrac{3-n}{n-1}\right)}$, then $\tilde{F}=\rho^{\left(\dfrac{3-n}{n-1}\right)}\left(\dfrac{\ddot{\rho}}{\rho}+\left(\dfrac{2-n}{n-1}\right)\left(\dfrac{\dot{\rho}}{\rho}\right)^{2}+(2\sigma^{2}+\tilde{R})\right)$ satisfies all the Helmholtz conditions provided $(2\sigma^{2}+\tilde{R})$ is a function of $\rho$ only. A Lagrangian can be constructed as follows
\begin{equation}
	L=\dfrac{1}{2}\rho^{\left(\dfrac{4-2n}{n-1}\right)}\dot{\rho}^{2}-V[\rho]
\end{equation}
Here, $V$ is the classical potential corresponding to the dynamical system representing the congruence. If we assume $V$ to be a function of $\rho$ only i.e, $V[\rho]=V(\rho)$ then the Euler-Lagrange equation $\dfrac{\partial L}{\partial \rho}=\dfrac{d}{d\Lambda}\left(\dfrac{\partial L}{\partial \dot{\rho}}\right)$ gives back the RE (\ref{eq34}). Now we switch to Hamiltonian formulation. The canonically conjugate momentum corresponding to $\rho$ is $\Pi_{\rho}=\dfrac{\partial L}{\partial \dot{\rho}}=\rho^{\left(\dfrac{3-n}{n-1}\right)}$ (using $\dot{\rho}=\rho \Theta$). The Hamiltonian is given by
\begin{equation}
	H=\dfrac{1}{2}\rho^{\left(\dfrac{2n-4}{n-1}\right)}\Pi_{\rho}^{2}+V[\rho]
\end{equation} 
The Hamilton's equation of motion are
\begin{eqnarray}
	\dot{\rho}=\rho \Theta\label{eq39}\\
	\rho^{\left(\dfrac{3-n}{n-1}\right)}\left(\dfrac{\ddot{\rho}}{\rho}+\left(\dfrac{2-n}{n-1}\right)\left(\dfrac{\dot{\rho}}{\rho}\right)^{2}+(2\sigma^{2}+\tilde{R})\right)=0\label{eq40}
\end{eqnarray} of which (\ref{eq39}) gives the definition of conjugate momentum while (\ref{eq40}) gives the RE.
\subsection{Canonical Quantization and Wheeler-DeWitt equation: Canonical approach}
We canonically quantize the system under consideration. For this $\rho$ and $\Pi_{\rho}$ are promoted to operators such that they satisfy the canonical commutation relation $[\hat{\rho},\hat{\Pi_{\rho}}]=i\hbar$ acting on the geometric flow state $\Psi[\rho,\Lambda]$. In $\rho$-representation, we have
\begin{equation}
	\hat{\rho}=\rho,~\hat{\Pi}=-i\hbar\dfrac{\partial}{\partial \rho}
\end{equation}
The operator version of the Hamiltonian is given by
\begin{equation}
	\hat{H}=-\dfrac{\hbar^{2}}{2}\rho^{\left(\dfrac{2n-4}{n-1}\right)}\dfrac{\partial^{2}}{\partial \rho^{2}}+V[\rho]\label{eq42}
\end{equation}
The evolution equation for the geodesic flow state $\psi$ can be written as
\begin{equation}
	\hat{H}\Psi=i\hbar\dfrac{\partial \Psi}{\partial \Lambda}
\end{equation}This can be interpreted as the evolution equation for quantized null geodesic congruence or the quantum analogue of the classical null RE for a general space-time of dimension $(n+1)$. Choosing the following operator ordering
\begin{equation}
	\hat{H}=-\dfrac{\hbar^{2}}{2}\rho^{\left(1-\dfrac{1}{n-1}\right)}\dfrac{\partial}{\partial \rho}~\rho^{\left(1-\dfrac{1}{n-1}\right)}\dfrac{\partial}{\partial \rho}+V[\rho]
\end{equation} and a transformation of variable $v=(n-1)\rho^{\dfrac{1}{n-1}}$, we can write
\begin{equation}
	\hat{H}=-\dfrac{\hbar^{2}}{2}\dfrac{\partial^{2}}{\partial v^{2}}+V[v]
	\end{equation} which is symmetric with the norm,
\begin{equation}
	\lVert \Psi(v) \rVert^{2}=\int_{0}^{\infty}dv~\Psi^{*}\Psi
\end{equation}
	$\lVert \Psi(v) \rVert^{2}$ is the probability distribution of the system.
In metric formulation of Einstein gravity, there is notion of Hamiltonian constraint and operator version of it acting on the wave function of the universe i.e, $\hat{H}\Psi=0$. The reason may be attributed to the fact that $\hat{H}$ generates infinitesimal
gauge transformations. As the physical states must be invariant under gauge transformations so they should be invariant under the action of the group member associated
to $\hat{H}$ (its exponential). It can alternatively demonstrated as follows: Classical $H=0$ on the constraint surface and hence quantum mechanically $\hat{H}\Psi=0$. In the quantization process due to
Dirac, the  physical quantum states of
the associated Hilbert Space must be annihilated by the
operator version of $H$. The WD equation using the quantum Hamiltonian operator in (\ref{eq42}) can be written as
\begin{equation}
	\dfrac{d^{2}\Psi}{d\rho^{2}}-\dfrac{2}{\hbar^{2}}\rho^{\left(\frac{4-2n}{n-1}\right)}V(\rho)\Psi(\rho)=0
\end{equation}
\textbf{Application to Schwarzschild BH:}\\
We consider the Schwarzschild metric given by
\begin{equation}
	ds^{2}=-\left(1-\dfrac{2M}{r}\right)dt^{2}+\left(1-\dfrac{2M}{r}\right)^{-1}dr^{2}+r^{2}d\Omega_{2}^{2}\label{eq47}
\end{equation} where $d\Omega_{2}^{2}=d\theta^{2}+\sin^{2}\theta d\phi^{2}$. In this case $\rho=\dfrac{r^{2}}{\sqrt{1-\dfrac{2M}{r}}}=\dfrac{r^{\frac{3}{2}}}{\sqrt{r-2M}}$  and hence $v=2\sqrt{\rho}$ (as $n=3$). Thus, at the physical singularity $r=0$ of the Schwarzschild BH, $\rho=v=0$. For $n=3$, the WD equation (\ref{eq47}) reduces to
\begin{equation}
		\dfrac{d^{2}\Psi}{d\rho^{2}}-\dfrac{2}{\hbar^{2}}\rho^{-1}V(\rho)\Psi(\rho)=0	\label{eq49}
\end{equation}
\textbf{Case-I: $V=V_{0}\rho^{m^{}}$ or $V=\frac{V_{0}}{4^{m}}v^{2m}$, $m$} is a real constant\\

The WD equation reduces to
\begin{equation}
	\dfrac{d^{2}\Psi}{dv^{2}}-\dfrac{2V_{0}'}{\hbar^{2}}v^{2m}\Psi(v)=0
	\end{equation} where $V_{0}'=\dfrac{V_{0}}{4^{m}}$. The general solution is given by 
\begin{equation}
	\Psi(v)=\left((2V_{0}')^{\frac{1}{4(1+m)}}(2\hbar(m+1))^{\frac{-1}{2(m+1)}}\sqrt{v}\right)~T
\end{equation} where $T=T_{1}+T_{2}$ and the expressions for $T_{1}$ and $T_{2}$ are respectively given by
\begin{eqnarray}
	T_{1}=c_{1}I_{\frac{-1}{2(m+1)}}\left(\dfrac{\sqrt{2V_{0}'}v^{1+m}}{(m+1)\hbar}\right)\Gamma\left(\dfrac{1+2m}{2(1+m)}\right)\\
	T_{2}=c_{2}(-1)^{\frac{1}{2(1+m)}}I_{\frac{1}{2(m+1)}}\left(\dfrac{\sqrt{2V_{0}'}v^{1+m}}{(n+1)\hbar}\right)\Gamma\left(\dfrac{3+2m}{2(1+m)}\right)
\end{eqnarray} where $c_{1}$ and $c_{2}$ are arbitrary constants. $I_{\nu}(z)$ is Bessel $I(\nu,z)$ function and $\Gamma$ represents the gamma function. It is to be noted that gamma function is undefined at zero and negative integers. So the choice of $m$ is restricted to $(2m+1)>0$ or $m>-\dfrac{1}{2}$. The above expression for the wave function shows that $\lVert \Psi(v) \rVert^{2}=0$ as $v\rightarrow0$ or as $r\rightarrow0$. Hence, the probability of having the physical singularity at $r=0$ is zero if the classical potential corresponding to the dynamical system representing the congruence is of the form $V=\frac{V_{0}}{4^{m}}v^{2m}$ with $m>-\dfrac{1}{2}$.\\

\textbf{CaseII: $V=V_{0}\exp(-\lambda\rho)$ or $V=V_{0}\exp(-\frac{\lambda}{4})v^{2}$, $\lambda$} is a real constant

The WD equation takes the form
\begin{equation}
	\dfrac{d^{2}\Psi}{dv^{2}}-\dfrac{2}{\hbar^{2}}V_{0}\exp\left(\frac{\lambda v^{2}}{4}\right)\Psi(v)=0
\end{equation}
The general solution or the wave function of the universe for this choice of potential is given by
\begin{equation}
	\Psi(v)=d_{1}I_{0}\left(\dfrac{\sqrt{2V_{0}}\sqrt{\exp(-\frac{\lambda}{4})v^{2}}}{\lambda \hbar}\right)+2d_{2}K_{0}\left(\dfrac{2\sqrt{2V_{0}}\sqrt{\exp(-\frac{\lambda}{4})v^{2}}}{\lambda \hbar}\right)
\end{equation} where $d_{1}$ and $d_{2}$ are arbitrary constants. $I_{\nu}(z)$ and $K_{\nu}(z)$ stand for Bessel $I(\nu,z)$ and $K(\nu,z)$ functions respectively. Clearly, $\lVert \Psi(v) \rVert^{2}\neq0$ at $v=0$. One may note that, for this exponential choice of the potential there is a finite probability at $r=0$. So the physical singularity can not be avoided in quantum description with this particular choice of the classical potential.\\

\textbf{Remark:}
 The above situation can be demonstrated in another way. For the power law choice of potential given by $V=V_{0}\rho^{m}$, as we approach to the physical singularity at $r=0$ or equivalently $\rho\rightarrow0$, there is basically no effect of the classical potential ($V\rightarrow0$ as $\rho\rightarrow0$) for $m>0$ and $V$ becomes infinite ($V\rightarrow \infty$) as $\rho\rightarrow0$ for $-\frac{1}{2}<m<0$. Also, by probabilistic approach we saw that there is no physical singularity for this choice of the potential with $m>-\frac{1}{2}$. On the other hand, for the exponential choice as we approach to the physical singularity at $r=0$ or equivalently $\rho=0$, the potential is non-vanishing and has some finite effect ($V\rightarrow V_{0}$ as $\rho\rightarrow0$). Again using the probabilistic approach in this case, we saw that singularity at $r=0$ can not be avoided. This shows that, non-zero finite classical potential (near the physical singularity) corresponding to the dynamical system representing the null geodesic congruence favors the formation of  singularity in Schwarzschild BH.
 \subsection{Bohmian formalism: Causal Interpretation}
   Quantization of classical geodesic flow, quantized version of null RE and WD formalism of the quantum system so constructed find an immediate application in the singularity analysis.  As a consequence, the effect of classical potential in forming or avoiding the physical singularity of Schwarzschild BH has been examined. This motivates us to examine the role of quantum potential towards the formation or avoidance of singularity of Schwarzschild BH. In order to include the role of quantum potential along with the classical potential, we need to replace the classical geodesics by quantal Bohmian trajectories. The quantum version of the Hamiltonian constraint is nothing but the Wheeler-DeWitt equation \cite{Pal:2016ysz}
\begin{equation}
	\hat{H}(\tilde{q}_{\beta}(\Lambda),\tilde{p}^{\beta}(\Lambda))\Psi(q_{\beta})=0\label{eq56}
\end{equation}
 where $p^{\alpha}(\Lambda)$ and $q_{\alpha}(\Lambda)$ are the homogeneous degree of freedom obtained from the induced metric $\eta_{ab}$ and the conjugate momenta $\Pi^{ab}$.
   WKB approximation gives the transition from quantum solutions to the classical regime and the wave function is written in the form \cite{Chakraborty:2001za}
   \begin{equation}
   	\Psi=\exp\left(\dfrac{i}{\hbar}S\right)
   \end{equation} where $S$ has the power series expansion in $\hbar$ as follows:
\begin{equation}
	S=S_{0}+\hbar S_{1}+\hbar^{2}S_{2}+...\label{eq58}
\end{equation} One may recover the classical solution upon construction of a wave packet from $S_{0}$ as
\begin{equation}
	\Psi=\int\left(A(\vec{k})\exp\left(\dfrac{i}{\hbar}S_{0}\right)\right)d\vec{k}
\end{equation} $\vec{k}$ being a parameter. Substituting the ansatz with $S$ from (\ref{eq58}) in the WD equation (\ref{eq49}) we get
\begin{equation}
	\left(\dfrac{dS_{0}}{d\rho}\right)^{2}=\dfrac{2V(\rho)}{\rho}
\end{equation} or
\begin{equation}
	S_{0}=\sqrt{2}\int \rho^{-\frac{1}{2}}\sqrt{v(\rho)}d\rho+k_{0}
\end{equation} where $k_{0}$ is the constant of integration. Thus one may construct a wave as
\begin{equation}
	\Psi{\rho}=\int A(\vec{k})\exp\left(\frac{i}{\hbar}S_{0}(\vec{k},\rho)\right)d\vec{k}
\end{equation} where $A(\vec{k})$ is a sharply peaked Gaussian function. $\dfrac{\partial S_{0}(\rho)}{\partial \vec{k}}=0$ in case of a constructive interference. This gives a relation between $\vec{k}$ and $\rho$ i.e, $\vec{k}=\vec k(\rho)$. So the wave function can be written as
\begin{equation}
	\Psi(\rho)=\int A(\vec{k}(\rho))\exp\left(\dfrac{i\sqrt{2}}{\hbar}\int \rho^{\frac{1}{2}}\sqrt{V(\rho)}d\rho+k_{0}\right)\dfrac{d\vec{k}}{d\rho}d\rho
\end{equation}
Using the similar ansatz for the WKB approximation into the causal interpretation we have
\begin{equation}
	\Psi(\rho)=A(\rho)\exp\left(\dfrac{i}{\hbar}S(\rho)\right)
\end{equation}
Using this ansatz in to the WD equation (\ref{eq56}) one gets the Hamilton-Jacobi equation as
\begin{equation}
	\dfrac{1}{2}\eta_{ab}q(\gamma)\dfrac{\partial S}{\partial q_{\alpha}}\dfrac{\partial S}{\partial q_{\beta}}+U(q_{\gamma})+W(q_{\gamma})=0
\end{equation} where
\begin{equation}
	W(q_{\gamma})=-\dfrac{1}{A}\eta_{ab}\dfrac{\partial^{2}A}{\partial q_{\alpha}\partial q_{\beta}}
\end{equation} is termed as the quantum potential, $\eta_{ab}$ is the reduction of the supermetric to the minisuperspace. $U(q_{\gamma})$ is the particularization of the scalar curvature density $(-q^{\frac{1}{2}}R)$ of the space-like hyper-surface. The trajectories $q_{\alpha}(\Lambda)$ due to causal interpretation must be real and observer independent \cite{Chakraborty:2001za}. The momentum corresponding to $q_{\alpha}$ can be obtained from the above Hamilton Jacobi equation as
\begin{equation}
	p^{\alpha}=\dfrac{\partial S}{\partial q^{\alpha}}
\end{equation} The usual momentum velocity relation is $p^{\alpha}=f^{\alpha\beta}\dfrac{\partial q_{\beta}}{\partial \Lambda}$. Thus equating the two ways of writing $p^{\alpha}$, one gets a system of first order differential equation
\begin{equation}
	\dfrac{\partial S}{\partial q_{\alpha}}=f^{\alpha\beta}\dfrac{\partial q_{\beta}}{\partial \Lambda}
\end{equation} These represent the quantal Bohmian trajectories. In the present quantum system the ansatz for the wave function is
\begin{equation}
	\Psi(\rho)=A(\rho)\exp\left(\dfrac{i}{\hbar}S(\rho)\right)
\end{equation}
Substituting this ansatz in the WD equation (\ref{eq49}) we get the Hamilton-Jacobi equation as
\begin{equation}
	\dfrac{1}{2\rho}\left(\dfrac{dS}{d\rho}\right)^{2}+V(\rho)+W(\rho)=0
\end{equation} where the quantum potential has the expression
\begin{equation}
	W(\rho)=-\dfrac{\hbar^{2}}{2\rho^{-1}}\dfrac{A''(\rho)}{A(\rho)}\label{eq}
\end{equation}
\textbf{Case-I: $A=A_{0}$}, a real constant\\
In this case the quantum potential is zero and the differential equation for quantum Bohmian trajectories for classical potential $V=V_{0}\rho^{m}$ is given by
\begin{equation}
	\dfrac{d\rho}{d\Lambda}=\sqrt{V_{0}'}\rho^{\frac{m+1}{2}}
\end{equation}
\begin{equation}
	\rho(\Lambda)=\rho_{0}\Lambda^{\frac{2}{1-m}}+l
\end{equation} where $\rho_{0}=\dfrac{\sqrt{V_{0}''}}{2}$, $V_{0}''=-2V_{0}$ and $l$ is a constant of integration. It is to be noted that, the trajectories pass through the physical singularity (characterized by $\rho=0$) in finite value of the affine parameter $\Lambda=\left(-\dfrac{l}{\rho_{0}}\right)^{\frac{1-m}{2}}$.

If we consider the choice $V=V_{0}\exp(-\Lambda\rho)$, then the differential equation for Bohmian trajectories is given by
\begin{equation}
		\dfrac{d\rho}{d\Lambda}=\sqrt{V_{0}}\rho^{\frac{1}{2}}\exp\left(-\dfrac{\lambda}{2}\rho\right)
\end{equation} The Bohmian trajectories are given by
\begin{equation}
	-\sqrt{\frac{2}{\mu}}\Gamma\left(\dfrac{1}{2},\dfrac{\mu\rho}{2}\right)=\sqrt{V_{0}'}\Lambda+\rho_{0}
\end{equation} where $\mu=-\lambda$ and $\rho_{0}$ is a constant of integration. In this choice of the classical potential also, the trajectories pass through the physical singularity in finite value of the affine parameter $\Lambda=-\left(\dfrac{\sqrt{\frac{2\pi}{\mu}}+\rho_{0}}{\sqrt{V_{0}'}}\right)$. This shows that the quantum trajectories behave classically for vanishing quantum potential and focus at the physical singularity within finite value of the affine parameter.\\

\textbf{Case-II.a: $A(\rho)=\rho^{L}$, $L\neq0,1$}\\
In this case the quantum potential is given by
\begin{equation}
	W(\rho)=\dfrac{\hbar^{2}L(1-L)}{2\rho}
\end{equation} If we consider power law choice of classical potential given by $V=V_{0}\rho^{m}$, then near the physical singularity $W(\rho)>>V(\rho)$ so that $W(\rho)-V(\rho)\approx W(\rho)$. So we neglect the classical potential in this case. The trajectories can be found as
\begin{equation}
	\rho(\Lambda)=\hbar\sqrt{L(L-1)}\Lambda+\rho_{0}
\end{equation}
This shows that, the trajectories will pass through the physical singularity in finite $\Lambda$ for proper fractional power law choice of the pre-factor $A$ present in the ansatz for the wave function.

\textbf{Case-II.b: $A=A_{0}\exp(-\alpha \rho)$}, $\alpha$ is a real constant\\
In this case the quantum potential is given by
\begin{equation}
	W(\rho)=-\dfrac{\hbar^{2}}{2}\alpha^{2}\rho
\end{equation} and for $V=V_{0}\rho^{m},~m=1$, the trajectories are given by
\begin{equation}
	\rho=\rho_{0}\exp(\sqrt{(\hbar^{2}\alpha^{2}-2V_{0})}\Lambda)
\end{equation} In this case, the quantal Bohmian trajectories will not pass through the physical singularity at $r=0$ of the Schwarzschild BH for any finite value of the affine parameter $\Lambda$. Thus, for non zero quantum potential (arising from exponential choice of the pre-factor present in the ansatz) and linear power form of classical potential with $V_{0}<\dfrac{\hbar^{2}\alpha^{2}}{2}$, the trajectories do not focus at the physical singularity of Schwarzschild BH. Further in this case, while approaching the physical singularity both classical as well as quantum potential are comparable.\\
\textbf{Note}: The case corresponding to $W=-\dfrac{\hbar^{2}}{2}\alpha^{2}\rho$ and $V=V_{0}\exp(-\lambda\rho)$ is not considered because if we approach the physical singularity, we get only the effect of classical potential but not the quantum potential which is not of our interest in the present section.
\section{Conclusion and discussion}
The paper focuses on Penrose's singularity theorem on black-holes. The notion of null geodesic incompleteness, existence of BH singularity are aligned with divergence of null Raychaudhuri matrix. The original Penrose's theorem has been restated in terms of a different geometric approach--matrix representation of tensor fields. In doing so, a matrix has been identified whose evolution essentially gives the matrix form of null RE. Thus, we have named it as \textit{null Raychaudhuri matrix}.\\

 The classical null RE and focusing theorem has been revisited. Since the paper is dedicated to black-holes so we have invoked null geodesic congruence and their corresponding evolution given by null RE. Further, with a suitable transformation of variable the null RE has been converted to a second order differential equation analogous to the evolution equation of a harmonic oscillator. Convergence condition has been stated using the frequency of the oscillator that varies with varying affine parameter. This approach has been further adopted to identify the physical singularity, coordinate singularity and event horizon of Schwarzschild BH. Reformulation of Penrose's theorem characterizes null geodesic incompleteness to the situation where the null Raychaudhuri matrix is undefined. Thus, one may interpret the mathematical analogue of null geodesic incompleteness with the divergence of null Raychaudhuri matrix. The notions of BH singularity, null geodesic incompleteness and undefined null Raychaudhuri matrix are all aligned. \\

The paper also shows two possible mitigation pathways of the physical singularity at $r=0$ of Schwarzschild BH---namely the canonical approach and causal approach. In canonical approach, we have deduced the evolution of a quantized null geodesic congruence and this turns out to be the quantum analogue of classical null RE. In this case, canonical quantization of the geodesic flow has been carried out followed by the formulation of WD equation. This has been done by the Lagrangian and Hamiltonian formulation of the RE. Subsequently, by considering the Schwarzschild metric the solution of the WD equation ($\Psi$, the wave function of the universe) so constructed has been found.  $\lVert \Psi \rVert^{2}$ gives the probability measure on minisuperspace or probability distribution of the system. In other way, $\Psi$ may be interpreted as the propagation amplitude of the congruence of null geodesics. If the wave packet so constructed is peaked along the classical solution, focusing of null geodesics may be avoided.\\

 We have found the wave function and $\lVert \Psi \rVert^{2}$ for two phenomenological choices of the classical potential, namely power law and exponential form in case of Schwarzschild BH. Using this probabilistic approach, we have found that the probability of having a physical singularity at ($r=0$) is zero if the classical potential assumes a power law form ($V=V_{0}\rho^{m}$) with power $m$ satisfying $m>-\frac{1}{2}$. While, it is not possible to mitigate the physical singularity at $r=0$ in case of exponential choice of the classical potential. Further, by examining the effect of classical potential near the physical singularity for both the choices we have inferred that, a non-zero finite classical potential (near the physical singularity) corresponding to the dynamical system representing the congruence of null geodesic favors the formation of singularity in Schwarzschild BH.\\

In causal approach, we have revisited the WKB approximation followed by Bohmian formalism. Subsequently, we have found the quantal Bohmian trajectories with or without quantum potential in case of Schwarzschild BH. It is a semi classical approach where the classical geodesics are replaced by the quantum trajectories. The trajectories are real and observer independent. The quantum effects are carried by quantum potential which appear in the Hamilton Jacobi equation along with the classical potential thereby making this approach semi classical. The trajectories have been found to behave classically in the absence of quantum potential and they pass through the physical singularity in finite affine parameter value. While, an exponential choice of the pre-factor present in the ansatz for the wave function gives rise to non-zero quantum potential and the trajectories do not pass through the physical singularity at $r=0$ for linear power of classical potential. Thus, the paper is an attempt to reformulate Penrose's theorem on BH singularity using the notion of null Raychaudhuri matrix and suggests some probable resolutions of Schwarzschild BH singularity via quantum description. However, it will be interesting to deal with Reissner–Nordström metric to see the effect of charge in formation or avoidance of singularity in future work.
\newpage
	\section*{Acknowledgment}
The authors thank the anonymous reviewer for insightful comments and suggestions which improved the quality and visibility of the paper. M.C thanks University Grants Commission (UGC) for providing the Senior Research Fellowship (ID:211610035684/JOINT CSIR-UGC NET JUNE-2021). S.C thanks FIST program of DST, Department of Mathematics, JU (SR/FST/MS-II/2021/101(C)). The authors are thankful to Inter University Centre for Astronomy and Astrophysics (IUCAA), Pune, India for their warm hospitality and research facilities where this work was carried out under the ``Visiting Research Associates" program of S.C.

	\end{document}